\documentclass[prd,aps,showpacs,preprintnumbers,amsmath,amssymb,11pt]{revtex4}
\usepackage{amsmath}
\usepackage{epsfig}
\usepackage{subfigure}
\begin{document}
\newcommand{\be}{\begin{equation}}\newcommand{\ee}{\end{equation}}
\newcommand{\bea}{\begin{eqnarray}}\newcommand{\eea}{\end{eqnarray}}
\newcommand{\bc}{\begin{center}}\newcommand{\ec}{\end{center}}
\def\no{\nonumber}
\def\eq#1{Eq. (\ref{#1})}\def\eqeq#1#2{Eqs. (\ref{#1}) and  (\ref{#2})}
\def\lsim{\raise0.3ex\hbox{$\;<$\kern-0.75em\raise-1.1ex\hbox{$\sim\;$}}}
\def\gsim{\raise0.3ex\hbox{$\;>$\kern-0.75em\raise-1.1ex\hbox{$\sim\;$}}}
\def\slash#1{\ooalign{\hfil/\hfil\crcr$#1$}}
\def\eff{\mbox{\tiny{eff}}}
\def\order#1{{\mathcal{O}}(#1)}
\def\pppm{B^0\to\pi^+\pi^-}
\def\pzpz{B^0\to\pi^0\pi^0}
\def\pppz{B^0\to\pi^+\pi^0}
\preprint{}
\title{Chiral Anomaly Effects and the  BaBar\\ Measurements of
the  $\gamma\gamma^{*}\to \pi^{0}$ Transition
Form Factor }
\author{T. N. Pham$^{a}$ and X. Y. Pham$^{b}$ }
\affiliation{
$^{a}$Centre de Physique Th\'{e}orique, CNRS,
Ecole Polytechnique, 91128 Palaiseau, Cedex, France\\
$^{b}$Laboratoire de Physique Th\'eorique et Hautes Energies, Paris,\\ 
Universit\'e Paris 6,  Unit\'e associ\'ee au CNRS, UMR 7589}

\date{\today}
\begin{abstract}
The recent BaBar measurements of the $\gamma\gamma^{*}\to \pi^{0}$
transition form factor  show  spectacular deviation from perturbative 
QCD prediction for large space-like  $Q^{2}$ up to 
$34\,\rm  GeV^{2}$. When plotted against $Q^{2}$, $Q^{2}F(Q^{2})$ 
shows steady increase with $Q^{2}$ in contrast with the flat $Q^{2}$ behavior
predicted by perturbative QCD, and at $34\,\rm  GeV^{2}$ is more 
than $50\%$ larger than the  QCD prediction. Stimulated by the BaBar 
measurements, we revisit our previous paper on the cancellation
 of anomaly effects in high energy processes $Z^{0}\to \pi^{0}\gamma$, 
$e^{+}e^{-}\to \pi^{0}\gamma$ and apply our results  
to the $\gamma^{*}\gamma\to \pi^{0}$ transition form factor measured 
in the $e^{+}e^{-}\to e^{+}e^{-}\pi^{0}$ process with one 
highly  virtual photon. We find that, the transition form factor 
$F(Q^{2})$ behaves as 
$(\frac{m^{2}}{Q^{2}})\times (\ln(Q^{2}/m^{2}))^{2}$ and produces a striking
agreement with  the BaBar data for $Q^{2}F(Q^{2})$ with $m=132\,\rm MeV$
 which also reproduces very well the CLEO data at lower
$Q^{2}$.
\end{abstract}
\maketitle


The  $\gamma^{*}\gamma\to \pi^{0}$ transition form factor at large
momentum transfer $Q^{2}$ which could
be  measured  in $Z^{0}\to \pi^{0}\gamma$ decay, in high energy 
$e^{+}e^{-}\to \pi^{0}\gamma$ or in  $e^{+}e^{-}\to e^{+}e^{-}\pi^{0}$
collisions \cite{Brodsky1} where one of the photon is highly 
virtual has been the subject of studies using the quark parton 
picture of hadrons for hard exclusive processes since the 
earlier days of  perturbative QCD. The interest in
this transition form factor lies in the fact that it is one of the simplest 
quantities to compute in QCD and relatively easy to measure. At
$Q^{2}=0$, it is given by the two-photon
$\pi^{0}$ decay  governed by the  Adler-Bell-Jackiw 
triangle chiral anomaly \cite{Adler,Adler1,Bell} which gives correctly 
the decay rate. At large $Q^{2}$, short-distance operator 
expansion(OPE) \cite{Frishman} or perturbative
QCD \cite{Lepage,Lepage1,Kroll} predicts $F(Q^{2})\sim 2\,f_{\pi}/Q^{2}$ (we use 
the convention $f_{\pi}=93\,\rm MeV$ in this paper). 
The earlier CLEO
data \cite{CLEO} give values for  $F(Q^{2})$ up to 
$Q^{2}= 8\,\rm GeV^{2}$ somewhat below 
the perturbative QCD(pQCD) prediction, though,  with  a possible rise 
for $Q^{2}\,F(Q^{2})$ above $2.5\,\rm GeV^{2}$ . Recently, the BaBar 
Collaboration has produced measurements  for the transition form factor from $4$ to
$34$  $\rm GeV^{2}$ \cite{Babar} which show spectacular deviation from
the perturbative QCD prediction as seen from the data for $Q^{2}\,F(Q^{2})$
which rise steadily with $Q^{2}$ in contrast with the rather
flat behavior predicted by pQCD and is more than $50\%$ above the QCD
prediction at $34\,\rm GeV^{2}$. There are also 
measurements by CELLO \cite{CELLO} up to $2.5\,\rm GeV^{2}$
which are shown in \cite{CLEO,Babar} As mentioned 
in \cite{Babar,Druzhinin}, recent calculations \cite{Bakulev,Bakulev1} using 
the light-cone sum rules method at next-to-leading order with various 
forms for the pion distribution amplitude, seem to obtain values 
for the transition form factor higher than the asymptotic limit
 of \cite{Lepage,Lepage1}, but with very different $Q^{2}$ behavior 
than the BaBar data for  $Q^{2}<15\,\rm GeV^{2} $ and are below 
the BaBar data for $Q^{2}$ in the range from $20$ to $40$ $\rm GeV^{2} $ 
\cite{Babar,Druzhinin}. A more recent work \cite{Wu} 
seems  to obtain  results consistent with the BaBar data for
 $Q^{2}>15\,\rm GeV^{2} $ with a very broad pion distribution
 amplitude. Most of these calculations 
use the short-distance OPE of Lepage-Brodsky \cite{Lepage,Lepage1}
to obtain the pion distribution amplitude, but, in the case of 
the $\gamma^{*}\to \gamma\pi^{0}$ transition form factor   
with one virtual photon with large $q^{2}$, the short-distance expansion
parameter $\omega = -2p\cdot q/q^{2}= 1$ ($p,q$ being the pion and 
virtual photon momentum, respectively), which is large and the
calculation of Lepage and Brodsky for fixed but large $Q^{2}$ cannot be 
trusted because the
corrections are important as mentioned in \cite{Manohar}.

 Without further questioning the validity of the
 perturbative QCD prediction \cite{Mikhailov,Kroll1,Agaev} which is 
based on the quark parton picture of the pion, one could already 
consider the role of chiral anomaly for  processes involving a
pion and highly virtual photons and the radiative decays of gauge 
bosons like $Z^{0}\to \pi^{0}\gamma$ or $W^{\pm}\to \pi^{\pm}\gamma$ 
decays \cite{Jacob}. In a previous paper \cite{Pham,Pham1}, we showed that,
for these processes, from the modified PCAC equation due to 
the Adler-Bell-Jackiw anomaly, the quark-parton contribution to 
the axial current divergence given by the triangle graph cancels the
anomaly term resulting in the suppression of the $Z^{0}\to
\pi^{0}\gamma$ decay amplitude as well as
the $\gamma^{*}\to \gamma\pi^{0}$ transition form factor measured in $e^{+}e^{-}\to
\pi^{0}\gamma$ . We also showed that for  $Q^{2}$ large as in 
$Z^{0}\to \pi^{0}\gamma$ decay or in $e^{+}e^{-}\to \pi^{0}\gamma$ process
with a highly virtual photon, the amplitude behaves like 
$(\ln(Q^{2}/m^{2}))^{2}/Q^{2}$, and  a similar behavior for space-like 
$Q^{2}$. The quantity $Q^{2}\,F(Q^{2})$ is found  to  rise with  
$Q^{2}$ as $(\ln(Q^{2}/m^{2}))^{2} $. This can be seen
from the $\ln(Q^{2}/m^{2})/Q^{2}$ behavior at large $Q^{2}$ of the 
absorptive part of the  triangle graph contribution to the 
divergence of the axial vector current
matrix element $<\pi^{0}|\partial_{\mu}A_{\mu}|\gamma^{*}\,\gamma>$. Hence
the $(\ln(Q^{2}/m^{2}))^{2}/Q^{2} $ behavior for the $\gamma^{*}\gamma\to
\pi^{0}$ transition form factor. 
This is one of the main differences between 
the chiral anomaly approach and the perturbative QCD quark parton 
approach to the $\gamma^{*}\gamma\to \pi^{0}$ transition form factor. In the chiral 
anomaly approach, the triangle graph gives us 
the absorptive and the dispersive part,  while the
quark parton approach based on short-distance operator expansion
 gives us only the real  part as given by  the tree graph 
at the lowest order in perturbative QCD. 

 Our calculation of the Adler-Bell-Jackiw  triangle
anomaly contribution to the $\gamma^{*}\gamma\to \pi^{0}$ transition
 form factor and  similar  calculations later on \cite{Bando,Kinoshita} 
are done at a time when few data on the transition form factors at 
large $Q^{2}$ are available \cite{CELLO,CLEO}. Now that the Babar data 
are available over a large range of large momentum transfer $Q^{2}$, 
it is relevant to compare data with the anomaly contribution,
considering the fact that, in pQCD the $\gamma^{*}\gamma\to \pi^{0}$ 
transition form factor depends on the pion distribution which is not known to a
good accuracy at present. For this reason, in this paper,  
we apply our previous analysis of the anomaly effects 
in $Z^{0}\to \pi^{0}\gamma$ to the 
$\gamma^{*}\gamma\to \pi^{0}$ transition form factor at space-like $Q^{2}$ . Using 
the generalized Goldberger-Treiman relation  to relate the 
quark mass  in the triangle graph to the quark-pion  Yukawa coupling $g$,
with  $m = g\,f_{\pi}$, as in the linear  sigma-model \cite{Cheng,Treiman}, we 
show that the BaBar data  can be reproduced with 
$m=132\,\rm MeV$ or $g=\sqrt{2}$, consistent with the 
$(\ln(Q^{2}/m^{2}))^{2}/Q^{2}$ behavior. 

  We  begin by  first recalling, for convenience,  our  derivation of the 
anomaly contribution to $Z^{0}\to\pi^{0}\gamma$ in \cite{Pham}.  
Similar to the $Z^{0}\to\pi^{0}\gamma$ decay, the 
 $\gamma^{*}\gamma\to \pi^{0}$ amplitude,
${\cal M} = <\pi^{0}(p)|T|\gamma^{*}(q)\gamma(k)> $
in the reaction $e^{+}e^{-}\to e^{+}e^{-}\pi^{0}$ has the form 
$\epsilon_{\mu}(q)\epsilon_{\nu}(k)N^{\mu\nu}(q,k) $ with:
\be
N^{\mu\nu}(q,k)= e^{2}F(q,k)Y^{\mu\nu}, \quad Y^{\mu\nu}= 
\epsilon^{\mu\nu\alpha\beta}q_{\alpha}k_{\beta}.
\label{amp}
\ee
where $p$ is the produced pion momentum in the final state ($p=q +k$),
$F(q,k)$ in the following, will be written as  $F(Q^{2})$, the form
factor in the kinematic region of the BaBar measurement, 
with  the virtual photon with momentum $q$ 
space-like ($q^{2}=-Q^{2}<0 $) while the  photon with momentum $k$
is almost on  the mass-shell ($k^{2}\simeq 0$) .
As with the derivation of the two-photon $\pi^{0}$ decay amplitude, we
start with the modified PCAC equation due to the Adler-Bell-Jackiw
triangle anomaly in  the presence of the electromagnetic
interactions. The divergence of the axial vector current associated 
with $\pi^{0}$  becomes:
\be
\partial_{\mu}A^{\mu}= f_{\pi}m_{\pi}^{2}\phi + S\frac{e^{2}}{16\pi^{2}}
\epsilon_{\alpha\beta\gamma\delta}F^{\alpha\beta}F^{\gamma\delta}
\label{dA}
\ee
with $F_{\mu\nu}$ the usual electromagnetic field strength tensor and $S$
are the sum of the squares of the charges and colors 
of quark contributing to the anomaly, which takes the value 
$S=1/2$ \cite{Adler}. Taking the matrix element of the l.h.s of
Eq. (\ref{dA}),
and seperating the $\pi^{0}$ pole term from the continuum, as previously
shown \cite{Pham}, we arrive at the expression for the 
$\gamma^{*}\gamma\to \pi^{0}$ amplitude:
\be
N^{\mu\nu} = \frac{1}{f_{\pi}}\left( p_{\tau}\tilde{R}^{\mu\nu\tau}(q,k)
- S\frac{e^{2}}{2\pi^{2}}Y^{\mu\nu}\right)
\label{Npi}
\ee
where $\tilde{R}^{\mu\nu\tau}(q,k) $ is 
the triangle graph  (the direct coupling between the three currents)
or the continuum contribution to the axial vector current matrix element
$<0|A_{\mu}|\gamma^{*}\gamma>$ defined as $R^{\mu\nu\tau}(q,k) $:
\be
R^{\mu\nu\tau}(q,k) = \tilde{R}^{\mu\nu\tau}(q,k)  - f_{\pi}
\frac{p^{\tau}N^{\mu\nu}(q,k)}{p^{2}-m_{\pi}^{2}}
\label{Rmnt}
\ee
 
As shown in \cite{Adler,Rosenberg}, gauge invariance and Bose symmetry
tell us that the divergence   $p_{\tau}R^{\mu\nu\tau}(q,k) $ is in
general proportional to $q^{2}$ and $k^{2}$ and does not vanish when one
or  both photons are off mass-shell. Only when
both photons are real($q^{2}= 0, k^{2}= 0$) that 
$p_{\tau}R^{\mu\nu\tau}(q,k) $ is  $O(p^{2})$ and becomes 
negligible. One can then apply Eq. (\ref{Npi}) to $\pi^{0}\to \gamma\gamma$
and finds that it is given by the anomaly \cite{Adler,Adler1,Bell}. For highly virtual
photon as in the present  $\gamma^{*}\gamma\to \pi^{0}$ transition form factor, we
will assume, as in \cite{Jacob}, that $\tilde{R}^{\mu\nu\tau}(q,k) $
is given by the triangle graph. From the expression given
 in \cite{Rosenberg}, we find, assuming equal mass for $u,d$ quarks in the
triangle graph:
\be
 p_{\tau}\tilde{R}^{\mu\nu\tau}(q,k) = e^{2}S\biggl(2mP(q,k) + \frac{1}{2\pi^{2}}\biggr)Y^{\mu\nu}
\label{Rt}
\ee 
where 
\be
 P(q,k) = \frac{m}{2\pi^{2}}\int_{0}^{1}dx\int_{0}^{1-x}\frac{dy}{D}
\label{Pm}
\ee 
and
\be
D= k^{2}y(1-y) + q^{2}x(1-x) -2q\cdot kxy -m^{2}
\label{D}
\ee 
The quark mass $m$ in the triangle graph is taken  as  a parameter 
to set the scale for the high energy limit, similar to the quark 
mass parameter used in the calculation of $e^{+}e^{-}\to q\bar{q}$ 
processes at high energy, where the asymptotic limit is reached 
when $Q^{2}\gg m^{2}$ and the quark-parton picture is valid and 
the $m^{2}/Q^{2}$ term in the cross section
 $\sigma(e^{+}e^{-}\to \rm hadrons)$ can be neglected.

When both photons are real ($q^{2}=0,k^{2}=0$), from  Eq. (\ref{Pm}), 
we get:
\be
2mP(q,k) =  -\frac{1}{2\pi^{2}} + O(p^{2})
\label{Pm0}
\ee 
which implies that the r.h.s of Eq. (\ref{Rt}) is $O(p^{2})$
in agreement with our previous remark that
$p_{\tau}\tilde{R}^{\mu\nu\tau}(q,k) = O(p^{2})  $. For our transition form factor
with time-like virtual photon $Q^{2}=s$, with $s> 4m^{2} $, we have, as
given in \cite{Pham}:
\be
2mP(q,k)=  \frac{1}{2\pi^{2}}\biggl(\frac{m^{2}}{s}\biggr)K(m^{2},s)
\label{Pms}
\ee
where 
\be
K(m^{2},s)=  \biggl(\ln\frac{1+\rho}{1-\rho} - i\pi\biggr)^{2} , \quad
\rho= \sqrt{1- 4m^{2}/s}, \quad s > 4m^{2}
\label{K}
\ee
For space-like $q$, with $q^{2} =-Q^{2}$ ($s= -Q^{2}$), with $s < 0$, 
by analytic continuation, the function $K(m^{2}, s)$ becomes real 
and is given by:
\be
K(m^{2},Q^{2})=  \biggl(\ln\frac{\rho +1}{\rho -1}\biggl)^{2} , \quad \rho= \sqrt{1+ 4m^{2}/Q^{2}}
\label{K1}
\ee
Using Eq. (\ref{Pms}) for $2mP(q,k)$ and Eq. (\ref{Rt}) for the divergence
$p_{\tau}\tilde{R}^{\mu\nu\tau}(q,k)$, we arrive at the final expression
for $\gamma^{*}\gamma\to \pi^{0}$ amplitude:
\be
N^{\mu\nu} = \frac{1}{f_{\pi}}\frac{e^{2}}{2\pi^{2}}S\biggl(\frac{m^{2}}{Q^{2}}K(m^{2},Q^{2})\biggr)Y^{\mu\nu}
\label{Nt}
\ee

 We note also that the term $2mP(q,k)$ in Eq. (\ref{Rt}) can be 
obtained directly from the triangle graph with the axial vector
 current replaced by the direct pion-quark vertex  with the  
$\gamma_{5}$ Yukawa coupling
$g=m/f_{\pi}$ as in the linear sigma model \cite{Cheng,Treiman} and
 PCAC is imposed on the pion-quark vertex . This is an  equivalent
 method to obtain the $\gamma^{*}\gamma\to \pi^{0}$ amplitude for 
large $Q^{2}$ without having to go  through  the proof of anomaly 
cancellation \cite{Pal}. The triangle graph with pion-quark vertex
 also gives us the term $2mP(q,k)$ for the $\pi^{0}\to \gamma\gamma$ 
decay amplitude, as shown in Eq. (\ref{Pm0}). 

For $S=1/2$, the scalar transition form factor $F(q,k)$ is 
given by:
\be
F(q,k) = \frac{1}{f_{\pi}}\frac{1}{4\pi^{2}}\frac{m^{2}}{s}K(m^{2},s)
\label{Ntf}
\ee
and for the space-like $\gamma^{*}\gamma\to \pi^{0}$ transition form factor, 
at large $Q^{2}\gg m^{2}$, the dominant term in Eq. (\ref{Ntf}) is
\be
F(Q^{2}) = \frac{1}{f_{\pi}}\frac{1}{4\pi^{2}}\frac{m^{2}}{Q^{2}}
\biggl(\ln{\frac{Q^{2}}{m^{2}}}\biggr)^{2}
\label{FF}
\ee
to be compared with the transition form factor for 
real photon(our normalization is the same as in \cite{Brodsky})
\be
F(q^{2}=0,k^{2}=0,p^{2}=0) = -\biggl(\frac{1}{4\pi^{2}f_{\pi}}\biggr)
\label{FF0}
\ee

We emphasize that our results Eqs. (\ref{Pms}--\ref{FF0}) are exact 
calculations and the result for the transition form factor given in 
Eq. (\ref{FF}) is valid for large $Q^{2}$ ( $Q^{2}\gg m^{2}$), including its 
$Q^{2}\to \infty$ limit. 

 This result, the $(\ln(Q^{2}/m^{2}))^{2} $ rise
for $Q^{2}\,F(Q^{2})$ which has been obtained   in  \cite{Pham}, are 
obtained later in \cite{Bando,Kinoshita}.  Ref. \cite{Bando}  obtains 
the anomaly contribution from the PCAC equation for the matrix 
element of the axial vector current between the photons, while  
Ref. \cite{Kinoshita} equates the divergence
of the axial vector current two-photon matrix element with its pion pole
terms. In our calculation, the quark mass parameter is taken as a
 dynamical(constituent) quark mass. The reason why we take constituent
 quark mass is also emphasized  in \cite{Bando,Kinoshita}. In fact  
 as argued in Ref. \cite{Bando}, because of the presence of
 the Nambu-Goldstone pion pole term in the PCAC equation, the 
divergence of the axial vector current should also  be non-vanishing 
in the limit of vanishing current quark mass, to cancel the longitudinal 
term generated by the pion pole term, as in the 
derivation of the Goldberger-Treiman relation for the pion-nucleon 
coupling constant

\begin{figure}[h]
\centering
\includegraphics[height=8.0cm,angle=0]{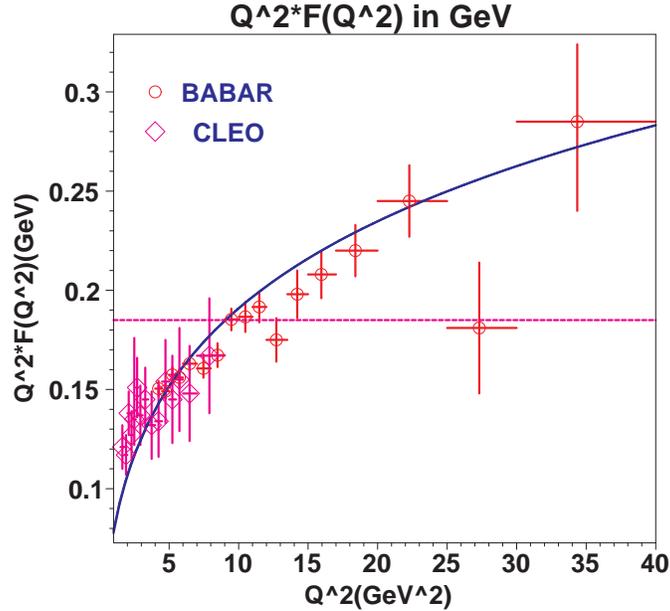}
\caption{Chiral anomaly prediction(solid line) for $Q^{2}F(Q^{2})$
compared with the BaBar and CLEO measured values  and the large $Q^{2}$ pQCD prediction (horizontal dash  line) of \cite{Lepage}
}
\label{fig1}
\end{figure}

As shown in Fig. (\ref{fig1}), our prediction for the quantity $Q^{2}F(Q^{2}) $
for $m=132\,\rm MeV$ fits very well the CLEO and BaBar data. The agreement 
with the BaBar data is striking, as our predicted transition form 
factor depends on only one parameter, the effective mass for quark 
in the triangle graph . We
note that recent works with various models for the pion distribution
mentioned above \cite{Bakulev,Wu} seem unable to obtain the 
rise of $Q^{2}F(Q^{2}) $ for $Q^{2}> 20\,\rm GeV^{2}$  as  the
BaBar measurements. Our value for the effective quark mass is 
consistent with the  high energy behavior of the 
$e^{+}e^{-}\to{\rm hadrons}$ cross section  for which
the ratio  $R=\sigma(e^{+}e^{-}\to {\rm hadrons})/\sigma(e^{+}e^{-}\to
\mu^{+}\mu^{-})$ is constant for $s$ above a few $\rm GeV^{2}$. 

 In conclusion, we have shown that chiral anomaly effects produce
a $(m^{2}/Q^{2})(\ln(Q^{2}/m^{2}))^{2}$ behavior for the
$\gamma \gamma^{*}\to \pi^{0}$ transition form factor at $Q^{2}\gg m^{2}$ in
 contrast with the 
$2f_{\pi}/Q^{2}$ behavior given by perturbative QCD. It is remarkable that
our simple expression for the transition form factor is able to explain the CLEO
and BaBar data.  Our prediction for the rise of $Q^{2}F(Q^{2})$ at higher
space-like $Q^{2}$ could be confirmed with further measurements.  
Similar behavior is expected for the time-like 
transition form factor and could be seen in $e^{+}e^{-}\to \pi^{0}\gamma$, as in  
$e^{+}e^{-}\to \eta^{(\prime)}\gamma$
for which the time-like transition form factor has been measured by CLEO
at $Q^{2}=14.2\,\rm GeV^{2} $ \cite{CLEO2} and is found to be close 
to the corresponding space-like values \cite{Druzhinin}, and by BaBar
at $Q^{2}=112\,\rm GeV^{2} $ \cite{Babar2}.
So, if chiral anomaly is indeed the cause of the rise like
$(\ln(Q^{2}/m^{2}))^{2}$  of $Q^{2}F(Q^{2})$, then, the pion might very well,
 for $\gamma\gamma^{*}\to \pi^{0}$ process at high energies,  behave like a 
Nambu-Goldstone boson, like the Adler zero in low-energy $\pi\pi$ 
scatterings and  in  $\psi^{\prime}\to J/\psi \pi\pi$  decay which are
obtained from chiral symmetry constrains.

\bigskip
{\em Note added}.  After the completion of this paper, we were informed
of the  papers by Dorokhov \cite{Dorokhov,Dorokhov1} in which similar
results are obtained using the triangle graph with  pion-quark
 coupling given in \cite{Ametller}.

\end{document}